\documentclass[%
 aps,
 prb,%
 amsmath,amssymb,
preprint,%
showpacs,
]{revtex4-1}

\usepackage{graphicx}
\usepackage{dcolumn}
\usepackage{bm}
\usepackage{hyperref}

\begin{document}

\title[Evanescent modes in PC]{Evanescent modes in Sonic Crystals: Complex dispersion relation and supercell approximation}

\author{V. Romero-Garc\'ia}
 \email{virogar1@mat.upv.es}
\author{J.V. S\'anchez-P\'erez}
 \affiliation{Centro de Tecnolog\'ias F\'isicas: Ac\'ustica, Universidad Polit\'ecnica de Valencia.}
\author{L.M. Garcia-Raffi}
 \affiliation{Instituto Universitario de Matem\'atica Pura y Aplicada, Universidad Polit\'ecnica de Valencia.}

\begin{abstract}
Evanescent modes in complete sonic crystals (SC) and SC with point
defects are reported both theoretically and experimentally in this
paper. Plane wave expansion (PWE) and, in general, $\omega(k)$
methods have been used to calculate band structures showing gaps
that have been interpreted as ranges of frequencies where no real
$k$ exists. In this work, we extend PWE to solve the complex
$k(\omega)$ problem applied to SC, introducing the supercell
approximation for studying one vacancy. Explicit matrix
formulation of the equations is given. This $k(\omega)$ method
enables the calculation of complex band structures, as well as
enabling an analysis of the propagating modes related with real
values of the function $k(\omega)$, and the evanescent modes
related with imaginary values of $k(\omega)$. This paper shows
theoretical results and experimental evidences of the evanescent
behavior of modes inside the SC band gap. Experimental data and
numerical results using the finite elements method are in very
good agreement with the predictions obtained using the $k(\omega)$
method.
\end{abstract}


\maketitle

\section{Introduction}
The propagation of scalar waves inside periodic structures has
been receiving growing interest in recent years. A great effort
has been made to understand the physics of these systems since the
acoustical properties of a periodic sculpture by Eusebio Sempere
were measured. \cite{Martinez05}

Phononic crystals (PCs) consist of an inhomogeneous periodic
distribution of elastic materials embedded in other elastic
materials with different properties. \cite{Kushwaha93, Sigalas93}
These systems are extensions of the photonic crystals
\cite{Yablonovitch, John} used for the propagation of elastic
waves through periodic elastic structures. If one of the elastic
materials is a fluid medium, then PCs are called sonic crystals
(SC). Several studies discuss the similarities and differences
between these periodic systems. \cite{Sigalas94, Economou93}

The periodicity of these systems is introduced in the solution of
the wave equation by means of Bloch's theorem. This solution leads
to the phenomenon of band gaps (BGs): frequency regimes where
waves do not propagate through the crystal. Traditionally, wave
propagation inside such systems was analyzed by means of the band
structures. Plane wave expansion (PWE) \cite{Kushwaha94PRB}
transforms the wave equation into an eigenvalue problem that can
be solved for each Bloch vector, $k$, in the irreducible first
Brillouin zone; and so obtaining the eigenfrequencies
$\omega(\vec{k})$ that constitute the band structures.  In the
case of SCs, it has been proven that  eigenfrequencies for an
arbitrary crystal structure and an arbitrary filling fraction
\cite{Halevi95} are real values.  A great number of applications
based on SCs are explained by the existence of BGs: acoustic
filters; \cite{Sanchez98} acoustic barriers; \cite{Sanchez02} or
wave guides. \cite{Khelif03, Khelif04}

Propagating waves inside a periodic media represent a set of
solutions to the wave equation that satisfy the translational
symmetry, and these are characterized by the transmission bands in
the PWE method. However, where the translational symmetry is
broken, finite periodic media or periodic media with point
defects,  can support the well known evanescent modes
characterized by a complex wave number, $k$.\cite{Joannopoulus08}
Recent experimental results \cite{Wu09} show measurements of the
sound levels recorded inside a point defect and behind an SC.
These authors observed that this level is higher inside the cavity
than behind the crystal. This fact clearly shows  both the
generation of a trapping mode (i.e. localized mode) inside the
point defect and its evanescent behavior outside the vacancy. Some
authors in the electromagnetic regime have measured the evanescent
modes in photonic crystals and revealed multi-exponential
decay.\cite{Engelen09}

Several extensions of the PWE method have been used to analyze the
propagation of sound through periodic systems in different
situations; for example, crystals with point defects have been
analyzed with PWE using the supercell approximation. \cite{Wu01,
Zhao09} The same methodology has been used to analyze the
influence of the following: constituent materials, plate
thickness, and the geometry of the array on the band structure in
two dimensional (2D) phononic crystal plates. \cite{Vasseur08}
However, these $\omega(\vec{k})$ methods interpret the BG as
frequency regimes where no real $k$ exists. Therefore, these
methods can only be used to study and characterize propagating
modes.

We have been motivated by the work of Hsue et al., \cite{Hsue05}
in which the PWE was extended for the case of photonic crystals to
calculate the complex $k$ in a 2D isotropic and in general 3D
anisotropic cases. In this paper we show the extended plane wave
expansion (EPWE) for the case of 2D SCs. The aim is to obtain the
band structures using the inverse expression $k(\omega)$, and with
a possibly complex $k$. Recent works show the calculation of
complex band structures for phononic crystals.\cite{Laude09,
Sainidou06} In the present work we show the explicit matrix
formulation and the approximation of supercell for analyzing the
complex relation dispersion of SCs. The extension of the
methodology enables us to characterize the evanescent and
propagating modes in complete SCs, as well as in SCs with point
defects.

In this paper we present novel measurements of the pressure in the
space between rows inside an SC. We have developed a 3D
computer-controlled automatic positioning system together with an
automatized acquisition system, called 3DReAMS (3D Robotized
e-Acoustic Measurement System). This system enables the pressure
field in trajectories inside a crystal to be measured, and we have
consequently analyzed the decay of the evanescent modes throughout
an SC. The imaginary part of the wave number of the evanescent
modes can be obtained experimentally with the measurements taken
by 3DReAMS. These data represent the experimental confirmation of
the analytical results obtained by the EPWE, as well as an
experimental analysis of propagating and evanescent modes in an
SC.

The paper is organized as follows. Section \ref{sec:PWE}
summarizes the main ingredients of the PWE for 2D SCs with the
explicit matrix formulation of the problem. In Section
\ref{sec:EPWE} we extend the PWE to the EPWE to solve the
eigenvalue problem $k(\omega)$. We show the matrix formulation, as
well as the EPWE, together with the supercell approximation for
studying the complex band structures of 2D SC with point defects.
In Section \ref{sec:results} the complex band structures of an SC
of PVC cylinders embedded in air are obtained with EPWE for a 2D
SC with, and without, point defects. Experimental results
validating the predictions of the EPWE for the evanescent and
propagating modes are shown in Section \ref{sec:experimental}.
Finally, the work is summarized in Section \ref{sec:Conclusions}.

\section{Plane wave method}
\label{sec:PWE}

Propagation of sound is described by the equation
\begin{eqnarray}
\frac{1}{\rho c^2} \frac{\partial^2 p}{\partial
t^2}=\nabla\left(\frac{1}{\rho}\nabla p \right)
\label{eq:acoustic}
\end{eqnarray}
where $c$ is the sound velocity, $\rho$ is the density of the
medium, and $p$ is the pressure.

In this paper we consider a system composed of an array of
straight, infinite cylinders made of an isotropic solid $A$,
embedded in an acoustic isotropic background $B$. There is
translational invariance in direction $z$ parallel to the
cylinders' axis; and the system has a 2D periodicity in the
transverse plane. By making use of this periodicity, we can expand
the properties of the medium in the Fourier series,
\begin{eqnarray}
\sigma=\frac{1}{\rho(\vec{r})}=\sum_{\vec{G}}\sigma_{\vec{k}}(\vec{G})e^{\imath \vec{G}\vec{r}} \label{eq:sigma},\\
\eta=\frac{1}{B
(\vec{r})}=\sum_{\vec{G}}\eta_{\vec{k}}(\vec{G})e^{\imath
\vec{G}\vec{r}}\label{eq:eta}.
\end{eqnarray}
$\vec{G}$ is the 2D reciprocal-lattice vector and
$B(\vec{r})=\rho(\vec{r})c(\vec{r})^2$ is the bulk modulus. For
the pressure $p$ we use the Bloch theorem and harmonic temporal
dependence,
\begin{eqnarray}
p(\vec{r},t)=e^{\imath (\vec{k}\vec{r}-\omega
t)}\sum_{\vec{G}}p_k(\vec{G})e^{\imath \vec{G}\vec{r}}.
\label{eq:pressure}
\end{eqnarray}

It is simple to show that \cite{Kushwaha94PRB}
\begin{eqnarray}
  \beta(\overrightarrow{G})= \left\{ \begin{array}{ll}
        \beta_{A}f+\beta_{B}(1-f)& \mbox{if $\overrightarrow{G} = \overrightarrow{0}$}\\
        \left(\beta_{A}-\beta_{B}\right)F(\overrightarrow{G}) & \mbox{if $\overrightarrow{G} \neq \overrightarrow{0}$}
        \end{array}\right.
\end{eqnarray}
where $\beta=(\sigma,\eta)$, and $F(\overrightarrow{G})$ is the
structure factor. For a circular cross section of radius $r$, the
structure factor is
\begin{eqnarray}
F(\overrightarrow{G})=\frac{1}{A_{uc}}\int_{A_{cyl}}
e^{{-i\overrightarrow{G}\overrightarrow{r}}}\overrightarrow{dr}=\frac{2f}{Gr}J_{1}(G).
\end{eqnarray}
$A_{uc}$ is the area of the unit cell, $A_{cyl}$ is the area of
the cylinder, and $J_1$ is the Bessel function of the first kind
of order $1$.

Using equations (\ref{eq:sigma}), (\ref{eq:eta}),
(\ref{eq:pressure}) and (\ref{eq:acoustic}) we
obtain\cite{Kushwaha94PRB}
\begin{eqnarray}
\sum_{\vec{G'}}\left((\vec{k}+\vec{G})\sigma_k(\vec{G}-\vec{G'})(\vec{k}+\vec{G'})-\omega^2\eta(\vec{G}-\vec{G'})\right)p_{\vec{k}}(\vec{G'})=0.
\label{eq:eigenproblem}
\end{eqnarray}
For $\vec{G}$ taking all the possible values, Equation
(\ref{eq:eigenproblem}) constitutes a set of linear, homogeneous
equations for the eigenvectors $p_{\vec{k}(\vec{G})}$ and
eigenfrequencies $\omega({\vec{k}})$. We obtain the band
structures when $\vec{k}$ scans the area of the irreducible region
of the first Brillouin zone.

Equation (\ref{eq:eigenproblem}) can be expressed by the matrix
formulation below
\begin{eqnarray}
\label{eq:matricial} \sum_{i=1}^3\Gamma_i\Sigma\Gamma_i P=\omega^2
\Omega P,
\end{eqnarray}
where i=1,2,3. The matrices $\Gamma_i$, $\Sigma$ and $\Omega$ are
defined as
\begin{eqnarray}
(\Gamma_i)_{mn}=\delta_{mn}(k_i+G_i^m).
\end{eqnarray}
The explicit matrix formulation is shown as follows:
\begin{eqnarray} \Gamma_i=\left(
\begin{array}{cccc}
k_i+G_i & 0 & \ldots & 0 \\
0 & k_i+G_i & \ldots & 0 \\
\vdots & \vdots & \ddots & \vdots\\
0 & \ldots & \ldots & k_i+G_i  \end{array}
\right)\label{eq:Gamma_matrix},\\[0.1cm]
\Sigma=\left( \begin{array}{ccc}
\sigma(\vec{G}_1-\vec{G}_1) & \ldots & \sigma(\vec{G}_1-\vec{G}_{N\times N}) \\
\vdots & \ddots & \vdots \\
\sigma(\vec{G}_{N\times N}-\vec{G}_1) & \ldots & \sigma(\vec{G}_{N\times N}-\vec{G}_{N\times N})\\
\end{array}
\right),\label{eq:Sigma_matrix}\\[0.1 cm]
\Omega=\left( \begin{array}{ccc}
\eta(\vec{G}_1-\vec{G}_1) & \ldots & \eta(\vec{G}_1-\vec{G}_{N\times N}) \\
\vdots & \ddots & \vdots \\
\eta(\vec{G}_{N\times N}-\vec{G}_1) & \ldots & \eta(\vec{G}_{N\times N}-\vec{G}_{N\times N})\\
\end{array}
\right),\label{eq:eta_matrix}\\[0.1 cm]
P=\left(\begin{array}{c}
P(\vec{G}_1)\\
\vdots\\
P(\vec{G}_{N\times N})\\
\end{array}
\right),
\end{eqnarray}
where $\vec{G}=(G_x, G_y, G_x)$. To solve (\ref{eq:matricial}) we
must truncate the matrices. If we chose $m=n=(-M,\ldots,M)$, the
size of the previous matrices is $N\times N=(2M+1)\times (2M+1)$.
$N\times N$ is usually the number of plane waves used in the
calculation.

By solving the system given in (\ref{eq:matricial}) for each Bloch
vector in the irreducible area of the first Brillouin zone, we
obtain $N\times N$ eigenvalues, $\omega^2$, which can be used to
represent the band structures, $\omega(\vec{k})$.

\section{Extended Plane Wave Method}
\label{sec:EPWE} In the $\omega(\vec{k})$ formulation, the
existence of BG is indicated by the absence of bands in determined
ranges of frequencies. However, BG could be understood by means of
the evanescent behavior of the internal modes. This interpretation
was predicted by some authors\cite{Joannopoulus08} when
approximating the second band near the BG by expanding
$\omega(\vec{k})$ to powers of $k$ around the edge $k=\pi/a$,
being $a$ the lattice constant of the array. These authors claimed
that as the BG is traversed,  the exponential decay grows as the
frequency nears the center of the BG. At a given frequency
$\omega$ inside the BG, the evanescent wave is characterized by a
complex value of its wave number $\vec{k}(\omega)$ and which the
imaginary part characterizes as the exponential-like decay of the
mode. In this section, we extend the previous PWE to the EPWE to
obtain $\vec{k}(\omega)$ and with a possibly imaginary $k$.

From Equation (\ref{eq:matricial}) we define the next vector,
\begin{eqnarray}
\Phi_i=\Sigma\Gamma_iP.
\end{eqnarray}
With this definition we can reformulate the eigenvalue problem
(\ref{eq:matricial}) as the equation system
\begin{eqnarray}
\Phi_i=\Sigma\Gamma_iP\nonumber\\
\omega^2\Omega P=\sum_{i=1}^3\Gamma_i\Phi_i.
\end{eqnarray}
To obtain an eigenvalue problem for $\vec{k}(\omega)$, we write
$\vec{k}=k\vec{\alpha}$, where $\vec{\alpha}$ is a unit vector.
Then (\ref{eq:Gamma_matrix}) can be written as
\begin{eqnarray}
\Gamma_i=\Gamma_i^0+k\alpha_iI,
\end{eqnarray}
where $I$ is the identity matrix, and
\begin{eqnarray}
\Gamma_i^0=\left(
\begin{array}{cccc}
G_i & 0 & \ldots & 0 \\
0 & G_i & \ldots & 0 \\
\vdots & \vdots & \ddots & \vdots\\
0 & \ldots & \ldots & G_i  \end{array}
\right),\label{eq:Gamma_matrix_b} \\[0.5cm]
 \alpha_i=\left(
\begin{array}{cccc}
\alpha_i & 0 & \ldots & 0 \\
0 & \alpha_i & \ldots & 0 \\
\vdots & \vdots & \ddots & \vdots\\
0 & \ldots & \ldots & \alpha_i  \end{array}
\right).\label{eq:alpha_matrix_b}
\end{eqnarray}

Equation (\ref{eq:matricial}) can then be written as
\begin{eqnarray}
\left(
\begin{array}{cc}
\omega^2\Omega -\sum_{i=1}^3\Gamma_i^0\Sigma\Gamma_i^0 & 0  \\
-\sum_{i=1}^3\Sigma \Gamma_i^0 & I\end{array} \right) \left(
\begin{array}{c}
P \\
\Phi' \end{array}\right)=k \left(
\begin{array}{cc}
\sum_{i=1}^3\Gamma_i^0\Sigma\alpha_i & I  \\
\sum_{i=1}^3\Sigma \alpha_i & 0\end{array} \right) \left(
\begin{array}{c}
P\\
\Phi'\end{array} \right) \label{eq:matricial_complex}
\end{eqnarray}
where $\Phi'=\sum_{i=1}^3\alpha_i\Phi_i$.

Equation (\ref{eq:matricial_complex}) represents a generalized
eigenvalue problem with $2N$ eigenvalues $k$, and possibly complex
numbers for each frequency. Complex band structures have been
calculated for the incidence direction characterized by  vector
$\vec{\alpha}$ by solving the previous eigenvalue equation for a
discrete number of frequencies and then sorting by continuity of
$k$. In contrast to the $\omega(\vec{k})$ method, the periodicity
is not relevant in this formulation of the problem  and
$k(\omega)$ does not follow the first Brillouin zone.

Because of the periodicity of the system, Bloch waves can be
expanded in a series of harmonics where each harmonic corresponds
with a value of $k$,  if $k$ is then a complex number, the
evanescent behavior of a wave with a predetermined frequency would
be multiexponential.\cite{Engelen09} The complex band structures
show the values of all of the complex values of $k$ which
contribute to the multi-exponential decay of the mode in the BG.
As we will see later, for the case of the SC analyzed in this
paper, we can only approximate the evanescent behavior in the
modes inside the BG  by considering the first term of this
harmonic expansion in terms of $k$.

\subsection{Supercell approximation}
One particularly interesting aspect of SCs is the possibility of
creating point defects that confine acoustic waves in localized
modes. \cite{Sigalas98, Zhao09} Because of the locally breaking
periodicity of the structure, defect modes can be created within
the BG. These defect modes are strongly localized around the point
defect: once the wave is inside the defect, it is trapped because
the borders of the defect act as perfect mirrors for waves with
frequencies in the BG. Localization depends on several parameters,
such as the size of the point defect. However, in finite periodic
structures the strength of sound localization also depends on the
size of the structure around the defect because of the exponential
decay of the outgoing wave.\cite{Wu09}

To analyze the propagation of waves inside periodic structures
with defects, authors have traditionally used PWE with supercell
approximation. The supercell method requires the lowest possible
interaction  between defects. This results in a periodic
arrangement of supercells that contain the point defect. With this
method we can obtain the relation $\omega(\vec{k})$ for crystals
with local defects and, for instance, the physics of wave guides
\cite{Khelif04, Vasseur08} or filters \cite{Khelif03} can be
explained.

In this section, we apply the supercell approximation to the EPWE.
This methodology enables us to obtain the relation $k(\omega)$ for
defect modes. It will be interesting to discover how the imaginary
part of the wave vector inside the BG changes with the creation of
the defect.

Consider an SC with primitive lattice vectors $\vec{a}_i$
($i=1,2,3$). The supercell is a cluster of $n_1\times n_2\times
n_3$ scatterers periodically placed in  space. The primitive
lattice vectors in the supercell approximation are
$\vec{a'}_i=n_i\vec{a}_i$, and the complete set of lattices in the
supercell approximation is $\{R'|R'=l_i\vec{a'}_i\}$, where $n_i$
and $l_i$ are integers. The primitive reciprocal vectors are then
\begin{eqnarray}
\vec{b'}_i=2\pi \frac{\varepsilon_{ijk}\vec{a'}_j\times
\vec{a'}_k}{\vec{a'}_1\cdot(\vec{a'}_2\times \vec{a'}_3)}
\end{eqnarray}
where $\varepsilon_{ijk}$ is the completely anti-symmetrical
three-dimensional Levi-Civita symbol. The complete set of
reciprocal lattice vectors in the supercell is
$\{\vec{G}|\vec{G}_i=N_i\vec{b'}_i\}$ where $N_i$ are integers.

Finally,  the structural factor of the supercell in this
approximation has to be computed while taking into account the
size of the supercell. If we consider a 2D SC with cylindrical
scatterers with a radius $r$ and an $n_1\times n_2$ sized
supercell, the structure factor of the supercell is expressed by
\begin{eqnarray}
F(\vec{G})=\sum_{i=-(n_1-1)/2}^{(n_1-1)/2}\sum_{j=-(n_2-1)/2}^{(n_2-1)/2}e^{\imath(ia|\vec{G}_1|+ja|\vec{G}_2|)P(\vec{G})}
\end{eqnarray}
where
\begin{eqnarray}
P(\vec{G})=\frac{2f}{Gr}J_{1}(G).
\end{eqnarray}
$f$ is the filling fraction of the supercell, $G=|\vec{G}|$ and
$a$ is the lattice constant of the 2D periodic system.

By introducing the previous expressions in the matrices of the PWE
(\ref{eq:matricial}), or in the case of the EPWE
(\ref{eq:matricial_complex}), we can then use the supercell
approximation to calculate the band structure of a periodic
structure with, and without, a point defect.

\section{Numerical Results}
\label{sec:results} We consider a 2D SC consisting of PVC
cylinders of radius $r$ in an air background arranged in a square
lattice with a lattice constant $a$. The material parameters
employed in the calculations are $\rho_{air}=1.23$kg/$m^3$,
$\rho_{PVC}=1400$kg/$m^3$, $c_{air}=340$m/s and $c_{PVC}=2380$m/s.
We consider a filling fraction $f=\pi r^2/a^2\simeq0.65$. We have
used reduced magnitudes, \cite{Kushwaha94PRB} so the  reduced
frequency is $\Omega=wa/(2\pi c_{host})$, and the reduced wave
vector is $K=ka/(2\pi)$.

\subsection{Complete array}
\begin{figure}
\includegraphics[width=80mm,height=70mm,angle=0]{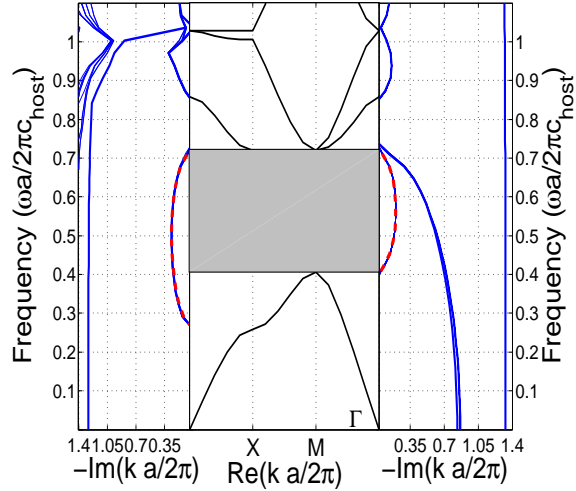}
\caption{\label{fig:complete}(Color online) Band structure of an
SC of PVC cylinders embedded in air with filling fraction
$f\simeq0.65$. The left panel represents the imaginary part of the
wave vector for each $\Gamma X$ direction frequency. The central
panel represents the real part of the wave vector, constrained in
the first Brillouin zone, for each frequency. The right panel
represents the imaginary part of the wave vector for each $\Gamma
M$ direction frequency. The red dashed line represents the
imaginary part of the wave vector of the evanescent modes inside
the BG. Reduced magnitudes have been used.}
\end{figure}

In Figure \ref{fig:complete} we can observe the complex band
structure obtained by EPWE for the SC described above. In the left
panel we have represented the imaginary part of the wave vector in
the $\Gamma X$ direction; in the right panel we have shown the
complex band structures in the $\Gamma M$ direction; and  the
central panel shows the real part of the band structures. The
imaginary part is not restricted in values of $k$; while the real
part is restricted to the first Brillouin zone. The area in gray
represents the full BG ranged between the frequencies
$\Omega_1=\omega_1 a/(2\pi c_{host})=0.4057$ and
$\Omega_2=\omega_2 a/(2\pi c_{host})=0.7189$. Note that the real
part of the complex band structures has exactly the same values as
in the case of the PWE.

In Figure \ref{fig:complete} we can observe that modes inside the
BG present purely imaginary wave vectors and these can be
characterized as evanescent modes with an exponential-like decay.
The elegant and intuitive explanation of the evanescent behavior
of modes inside the BG given by Joannopoulus\cite{Joannopoulus08}
is reproduced in  Figure \ref{fig:complete} in $\Gamma X$; as well
as in $\Gamma M$ directions (red dashed lines). The imaginary part
of the wave number for frequencies inside the BG grows with values
of frequency closer to the center of the BG; and disappears at the
edges of the BG. In other words, the rate of decay is greater for
frequencies closer to the center of the BG. We can also observe
that the imaginary part of the wave vector connects propagating
bands and so conserves the overall number of modes.

A recent paper has shown the multi-exponential decay of evanescent
modes in a photonic crystal.\cite{Engelen09} In Figure
\ref{fig:experimental}, we can observe clearly that each frequency
inside the BG is characterized by several values of $Im(k)$,
corresponding to the harmonics of the multi-exponential decay of
the evanescent modes. In the Section \ref{sec:results} we will see
that only the first value of the $Im(k)$ contributes to the decay
of the mode, and therefore higher harmonics can be neglected and
we can approximate in the same way as an exponential-like decay.

\subsection{Defect modes}

In this paper, point defects have been created by removing
cylinders in an SC.  We have used the EPWE method with supercell
approximation to analyze the propagating and evanescent behavior
of modes in an SC with point defects.

\begin{figure}
\includegraphics[width=80mm,height=70mm,angle=0]{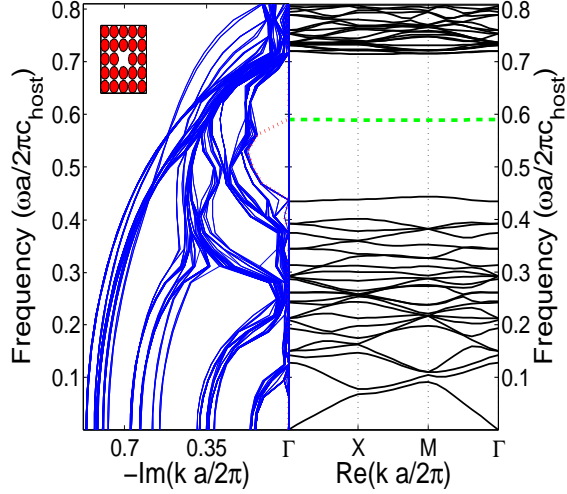}%
\caption{\label{fig:defect}(Color online) Band structure for an SC
with an internal defect, calculated using the EPWE with supercell
approximation. The left panel represents the imaginary part of the
wave vector for each  $\Gamma X$ direction frequency. The right
panel represents the real part, constrained in the first Brillouin
zone, of the wave vector for each frequency. The green dashed line
represents the frequency of the localized mode in the defect. The
red dotted line represents the imaginary part of the wave vector
of the evanescent modes inside the BG. Reduced magnitudes have
been used.}
\end{figure}

Figure \ref{fig:defect} shows the complex band structures for the
$\Gamma X$ direction and real band structures for an SC with a
point defect. In our case, we use only one direction of incidence
to analyze the complex band structure because the localized mode
appears at the same frequency for all the incidence directions.
The supercell used for the calculations is shown in the inset of
Figure \ref{fig:defect}. We can observe that the localized mode
appears at $\Omega_3=\omega_3 a/(2\pi c_{host})=0.59$ (green
dashed line). For frequencies in the BG, the borders of the point
defect act as perfect mirrors and produce the localized mode in
this cavity. The complex value of the $k$ number for the modes
inside the BG can be obtained by EPWE and becomes a purely real
value for the localized mode (red dotted line and green dashed
line). The value exactly coincides with the value obtained by PWE
with supercell approximation.

\section{Experimental results}
\label{sec:experimental}

We performed the experiments in an echo-free chamber sized
$8\times 6\times 3$m$^3$. To obtain the experimental dependence of
the pressure all along the SC, we measured the pressure field at
several points between two rows of the SC. To achieve this we
built a finite SC and placed the microphone inside the periodic
structure in a space between two rows. The finite 2D SC used in
this paper was made of PVC cylinders hung in a frame and measuring
5$a\times$5$a$. The radius of the cylinders was $r=10$cm, and the
lattice constant of the SC was $a=22$cm. With these parameters,
the finite SC has the same filling fraction ($f\simeq0.65$) as in
Section \ref{sec:results}, and the dimensions are large enough for
the microphone to be placed between the rows. The microphone used
was a prepolarized free-field 1/2" Type $4189$ B\&K. The diameter
of the microphone was $1.32$cm, which is approximately $0.06a$,
and so a low level of influence over the pressure field measured
is expected.

The 3DReAMS system is capable of sweeping the microphone through a
3D grid of measuring points located at any trajectory inside the
echo-free chamber. The motion of the robot was controlled by an
NI-PCI 7334. We analyzed the absolute value of the sound pressure
between two rows of the SC by moving the microphone in steps of
$1$ cm.

In Section \ref{sec:results} we  analyzed the upper and lower
frequencies of the BG for an SC of PVC cylinders with the filling
fraction value as in our experimental set up. By considering the
corresponding values of the parameters of our experimental SC, we
can obtain the frequency range of the BG. In our case, the BG
appears between $627$Hz and $1111$Hz. To measure the propagation
of sound inside the SC, we analyzed two different frequencies, one
inside the BG and the other  in the first transmission band. The
frequencies were $920$Hz and $442$Hz, respectively.

\begin{figure}
\includegraphics[width=80mm,height=70mm,angle=0]{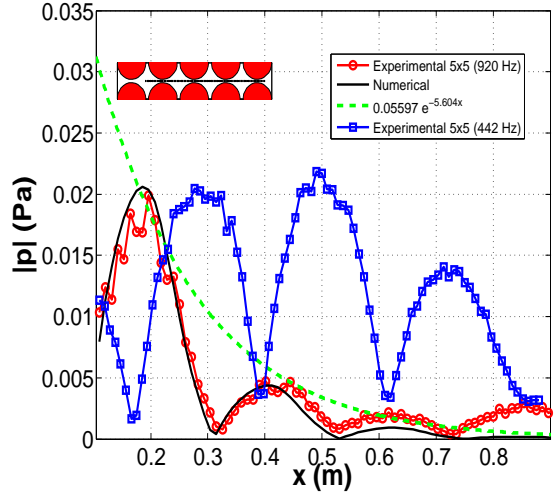}%
\caption{\label{fig:experimental}(Color online) Absolute value of
the pressure inside the SC in the positions between two rows. Blue
squares represent these values for a frequency outside of the BG,
$442$Hz. Red circles represent these values for a frequency inside
the BG, $920$Hz. Black dots represent the values used to fit the
exponential decay. Green line represents the fit of the
exponential decay of the evanescent mode inside the BG. The black
continuous line represents the absolute values of the pressure
obtained by finite element methods.}
\end{figure}

In Figure \ref{fig:experimental} we show the experimental
measurements of the absolute value of the pressure inside SC for
propagating and evanescent modes. These experimental results
represents a novel measurement of the pressure field inside an SC.
The inset of Figure \ref{fig:experimental} shows the measured
points in steps of $1$ cm placed between two rows of cylinders
inside the SC using the 3DReAMS system. Blue squares with a
continuous blue polygonal line represent the absolute value of the
pressure of a frequency outside of the BG, that is $442$Hz. This
frequency represents a propagating mode inside the SC. Red circles
with a polygonal red continuous line represent the absolute value
of the pressure of a frequency inside the BG, that is $920$Hz. For
the last case, we can observe the decay of the pressure inside the
SC because of the evanescent behavior of the mode inside the BG.

In contrast to the propagating mode (blue squares with a blue
polygonal continuous line), the evanescent mode (red squares with
a red polygonal  continuous line) is practically extinguished at
the end of the crystal -- and just a small value remaining for the
emerging pressure. This characteristic of evanescent behavior in
finite SCs has been measured recently by Wu et al. \cite{Wu09} in
an SC with a point defect.

The value of the imaginary part of the first harmonic of the wave
vector for the $920$Hz frequency can be obtained from Figure
\ref{fig:complete}. Using the values of parameters of the SC, we
can observe a value $Im(k)=-5.6$m$^{-1}$. From experimental data
(see Figure \ref{fig:experimental}), we can fit the decay of the
evanescent mode. We have chosen the points with maximum values in
order to fit an exponential decay $ae^{bx}$. The values of the fit
are $a=0.05597\pm0.0103$Pa and $b=Im(k)=-5.60\pm1.45$m$^{-1}$.
Note that the experimental value is very close to the analytical
value, i.e., the assumption that only the first harmonic is needed
to represent the multiexponential decay of the evanescent mode is
correct.

By solving the scattering problem inside the SC by means of the
finite element method (FEM) we can analyze the evanescent behavior
of the modes inside the BG of an SC. We have studied numerically
the absolute value of the sound pressure between two rows of an
SC. Continuity boundary conditions in the walls of the cylinders
and the radiation condition at the borders of the numerical domain
have been considered in the simulation. The black continuous line
in Figure \ref{fig:experimental} represents the absolute values of
pressure obtained numerically inside the SC, considering an
incidence of a plane wave with a frequency of $920$Hz. The
correspondence between the experimental data (red polygonal line
with open red circles) and the numerical results is clear.

\section{Conclusions}
\label{sec:Conclusions} The propagation of waves inside periodic
structures consists of propagating and evanescent modes. $\omega
(\vec{k})$ methods can be used to analyze the propagating modes,
while evanescent modes are represented by the absence of $k$ for
some ranges of frequencies. In this paper, we extend the
$\omega(\vec{k})$ to the $k(\omega)$ method for the case of 2D
SCs. We present the formulation of the supercell approximation for
the $k(\omega)$ method. For the EPWE we have predicted the
evanescent nature of the modes inside the BG of an SC. In this
paper we have reported measurements of the exponential-like decay
of the acoustic field inside an SC. EPWE predicted a value for the
imaginary part of the first harmonic of the wave number,
$Im(k)=-5.6$m$^{-1}$; and by fitting an exponential decay,
$ae^{bx}$, the experimental value we have obtained is
$b=Im(k)=-5.60\pm1.45$m$^{-1}$. Therefore, we can conclude that
only the first harmonic contributes to the exponential-like decay
of the evanescent mode. We have also shown that the imaginary part
of the wave vector connects propagation bands and conserves the
overall number of modes.

We have also applied the EPWE with supercell aproximation to SC
with point defects. We have analyzed the case of one vacancy
observing the localized mode inside the BG predicted by EPWE. The
value of the $k$ number for this localized mode, that is purely
imaginary in the case of complete SC, changes to purely real and
it becomes in a passing mode as it was observed in the literature.
The frequency of the localized mode exactly coincides with the
value obtained by PWE.

Analytical, numerical, and experimental results reproduce with
very good agreement the complex values of the wave vector inside
the BG, meaning that these methodologies obtain good values for
the exponential-like decay of the evanescent modes in an SC. This
work shows the basis for the correct understanding of the design
of narrow filters and wave guides based on phononic or sonic
crystals with point defects.

\begin{acknowledgments}
The authors would like to thank Dr. E.A. S\'anchez-P\'erez for his
comments and suggestions and thank Daniel Fenollosa and Talleres
Ferriols for their help in  building the mechanical part of
3DReAMS. This work was supported by MEC (Spanish government) and
the European Regional Development Fund , under grants
MAT2009-09438 and MTM2009-14483-C02-02.
\end{acknowledgments}

%

\begin{thebibliography}{10}%
\makeatletter
\providecommand \@ifxundefined [1]{%
 \ifx #1\undefined \expandafter \@firstoftwo
 \else \expandafter \@secondoftwo
\fi
}%
\providecommand \@ifnum [1]{%
 \ifnum #1\expandafter \@firstoftwo
 \else \expandafter \@secondoftwo
\fi
}%
\providecommand \enquote [1]{``#1''}%
\providecommand \bibnamefont  [1]{#1}%
\providecommand \bibfnamefont [1]{#1}%
\providecommand \citenamefont [1]{#1}%
\providecommand\href[0]{\@sanitize\@href}%
\providecommand\@href[1]{\endgroup\@@startlink{#1}\endgroup\@@href}%
\providecommand\@@href[1]{#1\@@endlink}%
\providecommand \@sanitize [0]{\begingroup\catcode`\&12\catcode`\#12\relax}%
\@ifxundefined \pdfoutput {\@firstoftwo}{%
 \@ifnum{\z@=\pdfoutput}{\@firstoftwo}{\@secondoftwo}%
}{%
 \providecommand\@@startlink[1]{\leavevmode}%
 \providecommand\@@endlink[0]{}%
}{%
 \providecommand\@@startlink[1]{%
  \leavevmode
  \pdfstartlink
   attr{/Border[0 0 1 ]/H/I/C[0 1 1]}%
   user{/Subtype/Link/A<</Type/Action/S/URI/URI(#1)>>}%
  \relax
 }%
 \providecommand\@@endlink[0]{\pdfendlink}%
}%
\providecommand \url  [0]{\begingroup\@sanitize \@url }%
\providecommand \@url [1]{\endgroup\@href {#1}{\urlprefix}}%
\providecommand \urlprefix [0]{URL }%
\providecommand \Eprint[0]{\href }%
\@ifxundefined \urlstyle {%
  \providecommand \doi [1]{doi:\discretionary{}{}{}#1}%
}{%
  \providecommand \doi [0]{doi:\discretionary{}{}{}\begingroup
  \urlstyle{rm}\Url }%
}%
\providecommand \doibase [0]{http://dx.doi.org/}%
\providecommand \Doi[1]{\href{\doibase#1}}%
\providecommand \bibAnnote [3]{%
  \BibitemShut{#1}%
  \begin{quotation}\noindent
    \textsc{Key:}\ #2\\\textsc{Annotation:}\ #3%
  \end{quotation}%
}%
\providecommand \bibAnnoteFile [2]{%
  \IfFileExists{#2}{\bibAnnote {#1} {#2} {\input{#2}}}{}%
}%
\providecommand \typeout [0]{\immediate \write \m@ne }%
\providecommand \selectlanguage [0]{\@gobble}%
\providecommand \bibinfo [0]{\@secondoftwo}%
\providecommand \bibfield [0]{\@secondoftwo}%
\providecommand \translation [1]{[#1]}%
\providecommand \BibitemOpen[0]{}%
\providecommand \bibitemStop [0]{}%
\providecommand \bibitemNoStop [0]{.\EOS\space}%
\providecommand \EOS [0]{\spacefactor3000\relax}%
\providecommand \BibitemShut [1]{\csname bibitem#1\endcsname}%
\bibitem{Martinez05}%
  \BibitemOpen
  \bibfield{author}{%
  \bibinfo {author} {\bibfnamefont{R.}~\bibnamefont{Mart\'inez-Sala}}, \bibinfo
  {author} {\bibfnamefont{J.}~\bibnamefont{Sancho}}, \bibinfo {author}
  {\bibfnamefont{J.~V.}\ \bibnamefont{S\'anchez}}, \bibinfo {author}
  {\bibfnamefont{V.}~\bibnamefont{G\'omez}}, \bibinfo {author}
  {\bibfnamefont{J.}~\bibnamefont{Llinares}},\ and\ \bibinfo {author}
  {\bibfnamefont{F.}~\bibnamefont{Meseguer}},\ }%
  \bibfield{journal}{%
  \bibinfo {journal} {nature}\ }%
  \textbf{\bibinfo {volume} {378}},\ \bibinfo {pages} {241} (\bibinfo {year}
  {1995})%
  \bibAnnoteFile{NoStop}{Martinez05}%
\bibitem{Kushwaha93}%
  \BibitemOpen
  \bibfield{author}{%
  \bibinfo {author} {\bibfnamefont{M.~S.}~\bibnamefont{Kushwaha}}, \bibinfo
  {author} {\bibfnamefont{P.}~\bibnamefont{Halevi}}, \bibinfo {author}
  {\bibfnamefont{L.}~\bibnamefont{Dobrzynski}},\ and\ \bibinfo {author}
  {\bibfnamefont{B.}~\bibnamefont{Djafari-Rouhani}},\ }%
  \bibfield{journal}{%
  \bibinfo {journal} {Phys. Rev. Lett.}\ }%
  \textbf{\bibinfo {volume} {71}},\ \bibinfo {pages} {2022} (\bibinfo {year}
  {1993})%
  \bibAnnoteFile{NoStop}{Kushwaha93}%
\bibitem{Sigalas93}%
  \BibitemOpen
  \bibfield{author}{%
  \bibinfo {author} {\bibfnamefont{M.}~\bibnamefont{Sigalas}}\ and\ \bibinfo
  {author} {\bibfnamefont{E.}~\bibnamefont{Economou}},\ }%
  \bibfield{journal}{%
  \bibinfo {journal} {Solid State Commun.}\ }%
  \textbf{\bibinfo {volume} {86}},\ \bibinfo {pages} {141} (\bibinfo {year}
  {1993})%
  \bibAnnoteFile{NoStop}{Sigalas93}%
\bibitem{Yablonovitch}%
  \BibitemOpen
  \bibfield{author}{%
  \bibinfo {author} {\bibfnamefont{E.}~\bibnamefont{Yablonovitch}},\ }%
  \bibfield{journal}{%
  \bibinfo {journal} {Phys. Rev. Lett.}\ }%
  \textbf{\bibinfo {volume} {58}},\ \bibinfo {pages} {2059} (\bibinfo {year}
  {1987})%
  \bibAnnoteFile{NoStop}{Yablonovitch}%
\bibitem{John}%
  \BibitemOpen
  \bibfield{author}{%
  \bibinfo {author} {\bibfnamefont{S.}~\bibnamefont{John}},\ }%
  \bibfield{journal}{%
  \bibinfo {journal} {Phys. Rev. Lett.}\ }%
  \textbf{\bibinfo {volume} {58}},\ \bibinfo {pages} {2486} (\bibinfo {year}
  {1987})%
  \bibAnnoteFile{NoStop}{John}%
\bibitem{Sigalas94}%
  \BibitemOpen
  \bibfield{author}{%
  \bibinfo {author} {\bibfnamefont{M.~M.}~\bibnamefont{Sigalas}}, \bibinfo
  {author} {\bibfnamefont{E.~N.}~\bibnamefont{Economou}},\ and\ \bibinfo {author}
  {\bibfnamefont{M.}~\bibnamefont{Kafesaki}},\ }%
  \bibfield{journal}{%
  \bibinfo {journal} {Phys. Rev. B}\ }%
  \textbf{\bibinfo {volume} {50}},\ \bibinfo {pages} {3393} (\bibinfo {year}
  {1994})%
  \bibAnnoteFile{NoStop}{Sigalas94}%
\bibitem{Economou93}%
  \BibitemOpen
  \bibfield{author}{%
  \bibinfo {author} {\bibfnamefont{E.~N.}~\bibnamefont{Economou}}\ and\ \bibinfo
  {author} {\bibfnamefont{M.~M.}~\bibnamefont{Sigalas}},\ }%
  \bibfield{journal}{%
  \bibinfo {journal} {Phys. Rev. B}\ }%
  \textbf{\bibinfo {volume} {48}},\ \bibinfo {pages} {13434} (\bibinfo {year}
  {1993})%
  \bibAnnoteFile{NoStop}{Economou93}%
\bibitem{Kushwaha94PRB}%
  \BibitemOpen
  \bibfield{author}{%
  \bibinfo {author} {\bibfnamefont{M.~S.}~\bibnamefont{Kushwaha}}, \bibinfo
  {author} {\bibfnamefont{P.}~\bibnamefont{Halevi}}, \bibinfo {author}
  {\bibfnamefont{G.}~\bibnamefont{Mart\'inez}}, \bibinfo {author}
  {\bibfnamefont{L.}~\bibnamefont{Dobrzynski}},\ and\ \bibinfo {author}
  {\bibfnamefont{B.}~\bibnamefont{Djafari-Rouhani}},\ }%
  \bibfield{journal}{%
  \bibinfo {journal} {Phys. Rev. B}\ }%
  \textbf{\bibinfo {volume} {49}},\ \bibinfo {pages} {2313} (\bibinfo {year}
  {1994})%
  \bibAnnoteFile{NoStop}{Kushwaha94PRB}%
\bibitem{Halevi95}%
  \BibitemOpen
  \bibfield{author}{%
  \bibinfo {author} {\bibnamefont{H.}~\bibnamefont{Hern\'andez-Cocoletzi}}, \bibinfo {author}
  {\bibnamefont{A.}~\bibnamefont{Krokhin}},\ and\ \bibinfo {author} {\bibnamefont{P.}~\bibnamefont{Halevi}},\
  }%
  \bibfield{journal}{%
  \bibinfo {journal} {Phys. Rev. B}\ }%
  \textbf{\bibinfo {volume} {51}},\ \bibinfo {pages} {17181}
  (\bibinfo {year} {1995})%
  \bibAnnoteFile{NoStop}{Halevi95}%
\bibitem{Sanchez98}%
  \BibitemOpen
  \bibfield{author}{%
  \bibinfo {author} {\bibfnamefont{J.~V.}\ \bibnamefont{S\'anchez-P\'erez}},
  \bibinfo {author} {\bibfnamefont{D.}~\bibnamefont{Caballero}}, \bibinfo
  {author} {\bibfnamefont{R.}~\bibnamefont{M\'artinez-Sala}}, \bibinfo {author}
  {\bibfnamefont{C.}~\bibnamefont{Rubio}}, \bibinfo {author}
  {\bibfnamefont{J.}~\bibnamefont{S\'anchez-Dehesa}}, \bibinfo {author}
  {\bibfnamefont{F.}~\bibnamefont{Meseguer}}, \bibinfo {author}
  {\bibfnamefont{J.}~\bibnamefont{Llinares}},\ and\ \bibinfo {author}
  {\bibfnamefont{F.}~\bibnamefont{G\'alvez}},\ }%
  \bibfield{journal}{%
  \bibinfo {journal} {Phys. Rev. Lett.}\ }%
  \textbf{\bibinfo {volume} {80}},\ \bibinfo {pages} {5325} (\bibinfo {year}
  {1998})%
  \bibAnnoteFile{NoStop}{Sanchez98}%
\bibitem{Sanchez02}%
  \BibitemOpen
  \bibfield{author}{%
  \bibinfo {author} {\bibfnamefont{J.~V.}\ \bibnamefont{S\'anchez-P\'erez}},
  \bibinfo {author} {\bibfnamefont{C.}~\bibnamefont{Rubio}}, \bibinfo {author}
  {\bibfnamefont{R.}~\bibnamefont{Mart\'inez-Sala}}, \bibinfo {author}
  {\bibfnamefont{R.}~\bibnamefont{S\'anchez-Grandia}},\ and\ \bibinfo {author}
  {\bibfnamefont{V.}~\bibnamefont{G\'omez}},\ }%
  \bibfield{journal}{%
  \bibinfo {journal} {Appl. Phys. Lett.}\ }%
  \textbf{\bibinfo {volume} {81}},\ \bibinfo {pages} {5240} (\bibinfo {year}
  {2002})%
  \bibAnnoteFile{NoStop}{Sanchez02}%
\bibitem{Khelif03}%
  \BibitemOpen
  \bibfield{author}{%
  \bibinfo {author} {\bibfnamefont{A.}~\bibnamefont{Khelif}}, \bibinfo {author}
  {\bibfnamefont{A.}~\bibnamefont{Choujaa}}, \bibinfo {author}
  {\bibfnamefont{B.}~\bibnamefont{Djafari-Rouhani}}, \bibinfo {author}
  {\bibfnamefont{M.}~\bibnamefont{Wilm}}, \bibinfo {author}
  {\bibfnamefont{S.}~\bibnamefont{Ballandras}},\ and\ \bibinfo {author}
  {\bibfnamefont{V.}~\bibnamefont{Laude}},\ }%
  \bibfield{journal}{%
  \bibinfo {journal} {Phys. Rev. B}\ }%
  \textbf{\bibinfo {volume} {68}},\ \bibinfo {pages} {214301} (\bibinfo {year}
  {2003})%
  \bibAnnoteFile{NoStop}{Khelif03}%
\bibitem{Khelif04}%
  \BibitemOpen
  \bibfield{author}{%
  \bibinfo {author} {\bibfnamefont{A.}~\bibnamefont{Khelif}}, \bibinfo {author}
  {\bibfnamefont{M.}~\bibnamefont{Wilm}}, \bibinfo {author}
  {\bibfnamefont{V.}~\bibnamefont{Laude}}, \bibinfo {author}
  {\bibfnamefont{S.}~\bibnamefont{Ballandras}},\ and\ \bibinfo {author}
  {\bibfnamefont{B.}~\bibnamefont{Djafari-Rouhani}},\ }%
  \bibfield{journal}{%
  \bibinfo {journal} {Phys. Rev. E}\ }%
  \textbf{\bibinfo {volume} {69}},\ \bibinfo {pages} {067601} (\bibinfo {year}
  {2004})%
  \bibAnnoteFile{NoStop}{Khelif04}%
\bibitem{Wu09}%
  \BibitemOpen
  \bibfield{author}{%
  \bibinfo {author} {\bibfnamefont{L.}~\bibnamefont{Wu}}, \bibinfo {author}
  {\bibfnamefont{L.}~\bibnamefont{Chen}},\ and\ \bibinfo {author}
  {\bibfnamefont{C.}~\bibnamefont{Liu}},\ }%
  \bibfield{journal}{%
  \bibinfo {journal} {Physica B}\ }%
  \textbf{\bibinfo {volume} {404}},\ \bibinfo {pages} {1766} (\bibinfo {year}
  {2009})%
  \bibAnnoteFile{NoStop}{Wu09}%
\bibitem{Engelen09}%
  \BibitemOpen
  \bibfield{author}{%
  \bibinfo {author} {\bibfnamefont{R.}~\bibfnamefont{J.}~\bibfnamefont{P.}~\bibnamefont{Engelen}}, \bibinfo
  {author} {\bibfnamefont{D.}~\bibnamefont{Mori}}, \bibinfo {author}
  {\bibfnamefont{T.}~\bibnamefont{Baba}},\ and\ \bibinfo {author}
  {\bibfnamefont{L.}~\bibnamefont{Kuipers}},\ }%
  \bibfield{journal}{%
  \bibinfo {journal} {Phys. Rev. Lett.}\ }%
  \textbf{\bibinfo {volume} {102}},\ \bibinfo {pages} {023902} (\bibinfo {year}
  {2009})%
  \bibAnnoteFile{NoStop}{Engelen09}%
\bibitem{Wu01}%
  \BibitemOpen
  \bibfield{author}{%
  \bibinfo {author} {\bibfnamefont{F.}~\bibnamefont{Wu}}, \bibinfo {author}
  {\bibfnamefont{Z.}~\bibnamefont{Hou}}, \bibinfo {author}
  {\bibfnamefont{Z.}~\bibnamefont{Liu}},\ and\ \bibinfo {author}
  {\bibfnamefont{Y.}~\bibnamefont{Liu}},\ }%
  \bibfield{journal}{%
  \bibinfo {journal} {Phys. Lett. A}\ }%
  \textbf{\bibinfo {volume} {292}},\ \bibinfo {pages} {198} (\bibinfo {year}
  {2001})%
  \bibAnnoteFile{NoStop}{Wu01}%
\bibitem{Zhao09}%
  \BibitemOpen
  \bibfield{author}{%
  \bibinfo {author} {\bibfnamefont{Y.}~\bibnamefont{Zhao}}\ and\ \bibinfo
  {author} {\bibnamefont{L.B.Yuan}},\ }%
  \bibfield{journal}{%
  \bibinfo {journal} {J. Phys. D: Appl. Phys.}\ }%
  \textbf{\bibinfo {volume} {42}},\ \bibinfo {pages} {015403} (\bibinfo {year}
  {2009})%
  \bibAnnoteFile{NoStop}{Zhao09}%
\bibitem{Vasseur08}%
  \BibitemOpen
  \bibfield{author}{%
  \bibinfo {author} {\bibfnamefont{J.~O.}\ \bibnamefont{Vasseur}}, \bibinfo
  {author} {\bibfnamefont{P.~A.}\ \bibnamefont{Deymier}}, \bibinfo {author}
  {\bibfnamefont{B.}~\bibnamefont{Djafari-Rouhani}}, \bibinfo {author}
  {\bibfnamefont{Y.}~\bibnamefont{Pennec}},\ and\ \bibinfo {author}
  {\bibfnamefont{A.-C.}\ \bibnamefont{Hladky-Hennion}},\ }%
  \bibfield{journal}{%
  \bibinfo {journal} {Phys. Rev.B}\ }%
  \textbf{\bibinfo {volume} {77}},\ \bibinfo {pages} {085415} (\bibinfo {year}
  {2008})%
  \bibAnnoteFile{NoStop}{Vasseur08}%
\bibitem{Hsue05}%
  \BibitemOpen
  \bibfield{author}{%
  \bibinfo {author} {\bibfnamefont{Young-Chung}~\bibnamefont{Hsue}}, \bibinfo {author}
  {\bibfnamefont{Arthur J.}~\bibnamefont{Freeman}},\ and\ \bibinfo {author}
  {\bibfnamefont{Ben-Yuan}~\bibnamefont{Gu}},\ }%
  \bibfield{journal}{%
  \bibinfo {journal} {Phys. Rev B}\ }%
  \textbf{\bibinfo {volume} {72}},\ \bibinfo {pages} {195118} (\bibinfo {year}
  {2005})%
  \bibAnnoteFile{NoStop}{Hsue05}%
\bibitem{Laude09}%
  \BibitemOpen
  \bibfield{author}{%
  \bibinfo {author} {\bibfnamefont{V.}\ \bibnamefont{Laude}},
  \bibinfo {author} {\bibfnamefont{Y.}\ \bibnamefont{Achaoui}},
  \bibinfo {author} {\bibfnamefont{S.}\ \bibnamefont{Benchabane}},\ and\ \bibinfo {author}
  {\bibfnamefont{A.}\ \bibnamefont{Khelif}},\ }%
  \bibfield{journal}{%
  \bibinfo {journal} {Phys. Rev. B}\ }%
  \textbf{\bibinfo {volume} {80}},\ \bibinfo {pages} {092301} (\bibinfo {year}
  {2009})%
  \bibAnnoteFile{NoStop}{Laude09}%
\bibitem{Sainidou06}%
  \BibitemOpen
  \bibfield{author}{%
  \bibinfo {author} {\bibfnamefont{R.}\ \bibnamefont{Sainidou}},\
  and\  \bibinfo {author} {\bibfnamefont{N.}\ \bibnamefont{Stefanou}},\ }%
  \bibfield{journal}{%
  \bibinfo {journal} {Phys. Rev. B}\ }%
  \textbf{\bibinfo {volume} {73}},\ \bibinfo {pages} {184301} (\bibinfo {year}
  {2006})%
  \bibAnnoteFile{NoStop}{Sainidou06}%
\bibitem{Sigalas98}%
  \BibitemOpen
  \bibfield{author}{%
  \bibinfo {author} {\bibfnamefont{M.}~\bibnamefont{Sigalas}},\ }%
  \bibfield{journal}{%
  \bibinfo {journal} {J. Appl. Phys.}\ }%
  \textbf{\bibinfo {volume} {84}},\ \bibinfo {pages} {3026} (\bibinfo {year}
  {1998})%
  \bibAnnoteFile{NoStop}{Sigalas98}%
\bibitem{Joannopoulus08}%
  \BibitemOpen
  \bibfield{author}{%
  \bibinfo {author} {\bibfnamefont{J.~D.}\ \bibnamefont{Joannopoulus}},
  \bibinfo {author} {\bibfnamefont{S.~G.}\ \bibnamefont{Johnson}}, \bibinfo
  {author} {\bibfnamefont{J.~N.}\ \bibnamefont{Winn}},\ and\ \bibinfo {author}
  {\bibfnamefont{R.~D.}\ \bibnamefont{Meade}},\ }%
  \emph{\bibinfo {title} {Photonic Crystals. Molding the Flow of Light}}\
  (\bibinfo {publisher} {Princeton University Press},\ \bibinfo {year} {2008})%
  \bibAnnoteFile{NoStop}{Joannopoulus08}%
\end{thebibliography}
\end{document}